\documentclass[twocolumn]{revtex4}[12pt]
\usepackage{amsmath,empheq}
\usepackage{graphicx}
\usepackage{color}
\begin{document}
\title{Modulated dust-acoustic wave packets in an opposite polarity dusty plasma system}
\author{S. Jahan$^a$, N. A. Chowdhury, A. Mannan, and A. A. Mamun}
\address{Department of Physics, Jahangirnagar University, Savar,
Dhaka-1342, Bangladesh.\\
$^*$Email: jahan88phy@gmail.com$^a$}
\begin{abstract}
  The nonlinear propagation of the dust-acoustic bright and dark envelope solitons in an opposite
  polarity dusty plasma system (composed of non-extensive $q$-distributed electrons, isothermal ions,
  and positively as well as negatively charged warm dust) have been theoretically investigated.
  The reductive perturbation method (which is valid for a small, but finite amplitude limit)
  is employed to derive the nonlinear Schr\"{o}dinger equation. Two types of modes (namely, fast
  and slow dust-acoustic (DA) modes) have been observed. The conditions for the modulational
  instability (MI) and its growth rate in the unstable regime of the DA waves are significantly
  modified by the effects of non-extensive electrons, dust mass, temperatures of different
  plasma species, etc. The implications of the obtained results from our current investigation in
  space (viz. Jupiters magnetosphere, upper mesosphere, and comets tails) and laboratory (e.g. direct
  current and radio-frequency discharges, plasma processing reactors, fusion plasma devices, and solid-fuel
  combustion products, etc.) dusty plasmas are briefly discussed.
\end{abstract}
\maketitle
\section{Lead Paragraph}
\textbf{The nonlinear propagation of the dust-acoustic bright and dark envelope solitons in an opposite
polarity dusty plasma system (containing isothermal ions, non-extensive
$q$-distributed electrons, and positively as well as negatively charged warm dust) has been studied to
identify some new features (e.g. conditions for the modulational instabiliuty of the dust-acoustic
waves (DAWs), growth rate, formation of the bright and dark envelope solitons, etc.) of the DAWs. These new
features should be useful for understanding the nonlinear electrostatic disturbances in space and laboratory dusty plasmas.}
\section{Introduction}
\label{1sec:Introduction} Now-a-days, a great deal of interest has been devoted to
the new and fascinating field of dusty plasma due to their existence  in space
\cite{Shukla2002,Verheest1996a,Goertz1986,Whipple1985} and  massive applications in
laboratory plasmas \cite{Mamun1996a,Rao1990,Angelo1990}. The dusty plasma is generally
considered to be an ensemble of dust particles, free electrons, and ions, which has
been found widely in planetary rings, asteroid zones \cite{Rao1990}, cometary tails, magnetosphere as
well as the lower part of the earth's atmosphere \cite{Goertz1986,Whipple1985}.
However, the co-existence of opposite polarity dusty in plasmas, introduces a new dusty
plasma model called ``opposite polarity dusty plasma'' (OPDP) \cite{Mamun2016b}, whose
main species are positively and negatively charged warm massive dust \cite{Mamun2016b}. The exclusive property
of this OPDP, which makes it completely unique from other plasmas (viz. electron ion
and electron-positron plasmas) is that the ratio of the size of positively charged dust to that of negatively
charged dust can be smaller \cite{Chow1993} or larger \cite{Mendis1994a} or equal to unity
\cite{Mendis1994a}. Dust species can be charged either positively or negatively, because
of different charging processes, like photo-ionization, secondary electron emission, thermionic emission, and
collection of plasma particles (electrons and ions), etc. The existence of positively
charged dust has been observed in different regions of space (e.g. Jupiter's magnetosphere
\cite{Horányi1993a}, cometary tails \cite{Mendis1994a,Mendis1991b,Horányi1996b,Chow1993},
upper mesosphere \cite{Havnes1996}, etc). There are three main processes by which dust grains become
positively charged, namely, \cite{Fortov1998,Mamun2008c} (a) secondary emission of electrons from
the surface of the dust grains; (b) thermionic emission induced by the radiative heating; and (c)
photoemission in the presence of a flux of ultraviolet photons. There is also direct evidence
for the co-existence of both positively and negatively charged dust particles in different regions of space plasmas
(viz. cometary tails \cite{Horányi1996b,Mendis1991b}, upper mesosphere \cite{Havnes1996},
Jupiter's magnetosphere \cite{Horányi1996b,Horányi1993a,Mamun2008c,Chowdhury2017a}, etc.) and laboratory
devices (e.g. direct current and radio-frequency discharges, plasma processing reactors,
fusion plasma devices, solid-fuel combustion products, etc \cite{Shukla2002,Akhter2012}). In space and
astrophysical sectors, if the plasma particles move very fast compared to their thermal
velocities \cite{Renyi1955} then the Maxwellian distribution is no longer valid. For that reason, Tsallis
proposed the non-extensive statistics \cite{Tsallis1988,Driouch2017}, which is the
generalisation of Boltzmann-Gibbs-Shannon  entropy. The importance of Tsallis statistics lies with describing the
system of long range interactions, such as, plasmas and dusty plasmas \cite{Driouch2017}.

The researchers are focused on wave dynamics, specifically, dust-acoustic (DA) waves (DAWs),
dust-acoustic rogue waves (DARWs), and dust ion-acoustic waves (DIAWs) in understanding electrostatic
density perturbations and potential structures (viz. soliton, shock, vortices, and rogue waves
\cite{Chowdhury2017a,Dubinov2009,Gill2010,Tasnim2013}). The DAWs \cite{Rao1990} have low phase
velocity (where the inertia is provided by the dust mass and restoring force is provided by the thermal
pressure of electrons and ions \cite{Chowdhury2017a}) in comparison with the electron and ion thermal velocities.
The research on modulational instability (MI) of
DAWs in nonlinear and dispersive mediums have been increasing significantly due to their
existence in astrophysics, space physics \cite{Chowdhury2018b} as well as in application in many laboratory
situations \cite{Misra2006}. A great number of researchers have used the nonlinear
Schr\"{o}dinger (NLS) equation, which governs the dynamics of the DAWs and the formation of the envelope
solitons \cite{Misra2006}. Sayed and Mamun \cite{Sayed2007} studied the finite solitary potential structures
that exist in OPDP. El-Taibany \cite{El-Taibany2013} examined the DAWs in inhomogeneous four-component dusty
plasmas with opposite charge polarity dust grains and observed that only compressive soliton is created
corresponding to fast DAWs velocities. To the best knowledge of the authors, no attempt has been made on
MI, and corresponding dark and bright envelope solitons associated with the DAWs containing non-extensive
$q$-distributed electrons, isothermal ions, and positively and negatively charged warm massive dust. Therefore,
in our article, we will derive the NLS equation by employing the reductive perturbation method and examine
the conditions for the MI of the DAWs (in which inertia is provided by the dust mass and restoring force is
provided from the thermal pressure of non-extensive $q$-distributed electrons and isothermal ions).

The manuscript is organized as the following fashion: The governing equations of our considered plasma
model are stated in Sec. \ref{1sec:Model Equations}. The NLS equation is derived in Sec.
\ref{1sec:Derivation of the NLS equation}. The stability of DAWs and envelope solitons are examined in
Sec. \ref{1sec:Stability of DAWs} and \ref{1sec:Envelope solitons}, respectively. Finally, the summary
of our discussion is provided in Sec. \ref{1sec:Discussion}.
\section{Model Equations}
\label{1sec:Model Equations} In this paper, we consider a collisionless, fully ionized, unmagnetized four
component dusty plasma system formed of non-extensive q-distributed electrons (charge $-e$,
mass $m_e$), isothermal ions (charge $+e$; mass $m_i$) and inertial warm negatively charged dust particles
(charge $q_1=-z_1e$, mass $m_1$) as well as positively charged warm dust particles (charge
$q_2=+z_2e$; mass $m_2$); where $z_1$ ($z_2$) is the charge state of the negatively (positively) charged
warm dust particles. The negatively and positively charged warm dust particles can be displayed by continuity
and momentum equations, respectively, as given below:
\begin{eqnarray}
&&\hspace*{-2.3cm}\frac{\partial n_1}{\partial t}+\frac{\partial}{\partial x}(n_1 u_1)=0,
\label{1eq:1}\\
&&\hspace*{-2.3cm}\frac{\partial u_1}{\partial t} + u_1\frac{\partial u_1 }{\partial x}=\frac{z_1 e}{m_1}
\frac{\partial \varphi}{\partial x}-\frac{1}{m_1 n_1} \frac{\partial p_1}{\partial x},
\label{1eq:2}\\
&&\hspace*{-2.3cm}\frac{\partial n_2}{\partial t}+\frac{\partial}{\partial x}(n_2 u_2)=0,
\label{1eq:3}\\
&&\hspace*{-2.3cm}\frac{\partial u_2}{\partial t} + u_2\frac{\partial u_2 }{\partial x}=-\frac{z_2 e}{m_2}
\frac{\partial \varphi}{\partial x}-\frac{1}{m_2 n_2}\frac{\partial p_2}{\partial x},
\label{1eq:4}\
\end{eqnarray}
where $n_1(n_2)$ is the number densities of the negatively (positively) charged warm dust
particles; $t(x)$ is the time (space) variable; $u_1(u_2)$ is the fluid speed of the negatively
(positively) charged warm dust species; $e$ is the magnitude of the charge of the electron;
$\varphi$ is the electrostatic wave potential; and $p_1$ ($p_2$) is the adiabatic pressure of the negatively
(positively) charged warm dust particles. The system is closed through Poisson's equation, stated as
\begin{eqnarray}
&&\hspace*{-2.95cm}\frac{\partial^2\varphi}{\partial x^2}=4 \pi e (n_e-n_i+z_1 n_1-z_2 n_2),
\label{1eq:5}
\end{eqnarray}
where $n_i$ and $n_e$ are, respectively, the ion and electron number densities. The
quasi-neutrality condition at equilibrium can be shown as
\begin{eqnarray}
&&\hspace*{-4.35cm}n_{i0}+z_2 n_{20}=n_{e0}+z_1 n_{10},
\label{1eq:6}
\end{eqnarray}
where $n_{i0}$, $n_{20}$, $n_{e0}$, and $n_{10}$ are the equilibrium number densities of the
isothermal ions, positively charged warm dust particles, non-extensive $q$-distributed electrons,
and negatively charged warm dust particles, respectively. Now, in terms of normalized variables, namely,
$N_1=n_1/n_{10}$, $N_2=n_2/n_{20}$, $U_1=u_1/C_{d1}$ (with $C_{d1}$ being the sound speed of the
negatively charged warm dust particles), $U_2=u_2/C_{d1}$; $\phi=e\varphi/T_i$ (with $T_i$ being the
temperature of the isothermal ion); $T=t \omega_{pd1}$ (with $\omega_{pd1}$ being the plasma
frequency of the negatively charged warm dust particles), $X=x/\lambda_{Dd1}$ (with $\lambda_{Dd1}$
being the Debye length of the negatively charged warm dust species); $C_{d1}=(z_1T_i/m_1)^{1/2}$,
$\omega_{pd1}=(4 \pi e^2z_1^2 n_{10}/m_1)^{1/2}$; $\lambda_{Dd1}=(T_i/4 \pi e^2 z_1 n_{10})^{1/2}$;
$p_1=p_{10}(n_1/n_{10})^\gamma$ [with $p_{10}$ being the equilibrium adiabatic pressure of the
negatively charged warm dust particles and $\gamma=(N+2)/N$, where $N$ is the degree of freedom,
for one-dimensional case, $N=1$ so that $\gamma=3$], $p_{10}=n_{10}T_1$ (with $T_1$ being the
temperature of the negatively charged warm dust particles); $p_2=p_{20}(n_2/n_{20})^3$ (with $p_{20}$
being the equilibrium adiabatic pressure of the positively charged warm dust particles), and
$p_{20}=n_{20}T_2$ (with $T_2$ being the temperature of the positively charged warm dust particles).
After normalization, Eqs. (\ref{1eq:1})$-$(\ref{1eq:5}) becomes
\begin{eqnarray}
&&\hspace*{-1.7cm}\frac{\partial N_1}{\partial T}+\frac{\partial}{\partial X}(N_1 U_1)=0,
\label{1eq:7}\\
&&\hspace*{-1.7cm}\frac{\partial U_1}{\partial T} + U_1\frac{\partial U_1 }{\partial X} +
3\sigma_1  N_1\frac{\partial N_1 }{\partial X}=\frac{\partial \phi}{\partial X},
\label{1eq:8}\\
&&\hspace*{-1.7cm}\frac{\partial N_2}{\partial T}+\frac{\partial}{\partial X}(N_2 U_2)=0,
\label{1eq:9}\\
&&\hspace*{-1.7cm}\frac{\partial U_2}{\partial T} + U_2\frac{\partial U_2 }{\partial X}+
3\sigma_2  N_2\frac{\partial N_2 }{\partial X}=-\alpha \frac{\partial \phi}{\partial X},
\label{1eq:10}\\
&&\hspace*{-1.7cm}\frac{\partial^2 \phi}{\partial X^2}=(\mu_i+\beta-1)N_e-\mu_i N_i+N_1-\beta N_2,
\label{1eq:11}
\end{eqnarray}
where $\sigma_1=T_1/z_1 T_i$, $\sigma_2=m_1 T_2/z_1 m_2T_i $, $\alpha=m_1 z_2/m_2 z_1$, $\beta=z_2 n_{20}/z_1n_{10}$,
and $\mu_i=n_{i0}/z_1n_{10}$. The number densities of the non-extensive
q-distributed \cite{Chowdhury2017a} electron can be given by the following normalized equation
\begin{eqnarray}
&&\hspace*{-4.3cm}N_e=[1+(q-1)\delta \phi]^{\frac{
q+1}{2(q-1)}},
\label{1eq:12}
\end{eqnarray}
where  $\delta=T_i/T_e$ and $q$ is the non-extensive parameter describing the degree of
non-extensivity, i.e., $q=1$ indicates the Maxwellian distribution, whereas $q<1$ refers to the
super-extensivity, and the opposite condition $q>1$ corresponds to the sub-extensivity
\cite{Chowdhury2017c}. The number densities of the iso-thermally distributed \cite{Chowdhury2017a} ion can be
represented as
\begin{eqnarray}
&&\hspace*{-6.0cm}N_i= \mbox{exp}(-\phi).
\label{1eq:13}
\end{eqnarray}
Now, substituting Eqs. (\ref{1eq:12}) and (\ref{1eq:13}) into Eq. (\ref{1eq:11}), and extending
up to third order in $\phi$, we can write
\begin{eqnarray}
&&\hspace*{-0.90cm}\frac{\partial^2 \phi}{\partial X^2}+\beta N_2-N_1=\beta-1+
\gamma_1 \phi+\gamma_2\phi^2+\gamma_3 \phi^3+..,
\label{1eq:14}
\end{eqnarray}
where
\begin{eqnarray}
&&\hspace*{-0.8cm}\gamma_1=\frac{(\beta +\mu_i-1)(q+1)\delta}{2}+\mu_i,
\nonumber\\
&&\hspace*{-0.8cm}\gamma_2=\frac{(\beta +\mu_i-1)(q+1)(3-q)\delta^2}{8}-\frac{\mu_i}{2},
\nonumber\\
&&\hspace*{-0.8cm}\gamma_3=\frac{(\beta +\mu_i-1)(q+1)(q-3)(3q-5)\delta^3}{48}+\frac{\mu_i}{6}.
\nonumber\
\end{eqnarray}
The left hand side of Eq. (\ref{1eq:14}), is the contribution of electron and ion species.
\section{Derivation of the NLS equation}
\label{1sec:Derivation of the NLS equation} 
\begin{figure}[t!]
\centering
\includegraphics[width=70mm]{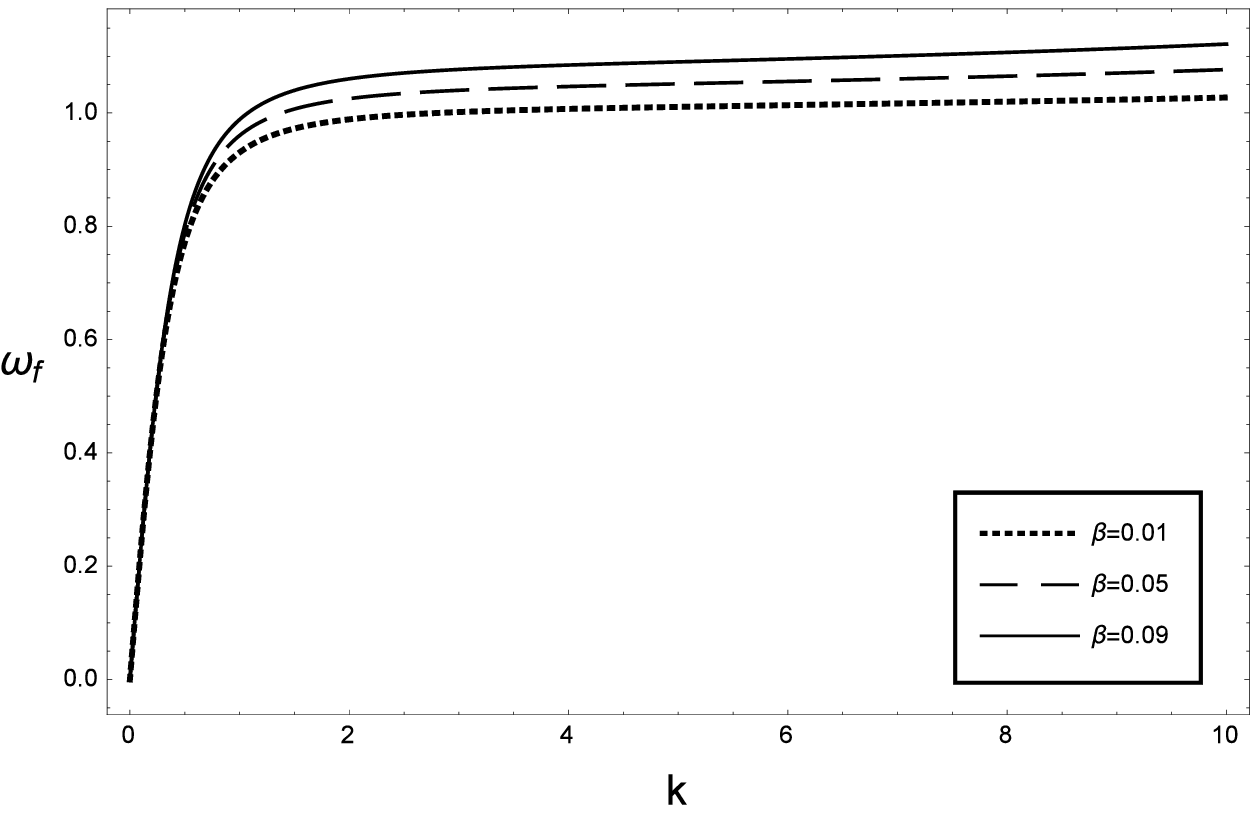}
\caption{The variation of $\omega_f$ with $k$ for different values of $\beta$;
along with $\alpha=2.0$, $\delta=0.3$, $\mu_i=0.4$, $\sigma_1=0.0001$, $\sigma_2=0.001$, and $q=1.5$.}
\label{1Fig:F1}
\vspace{0.8cm}
\includegraphics[width=70mm]{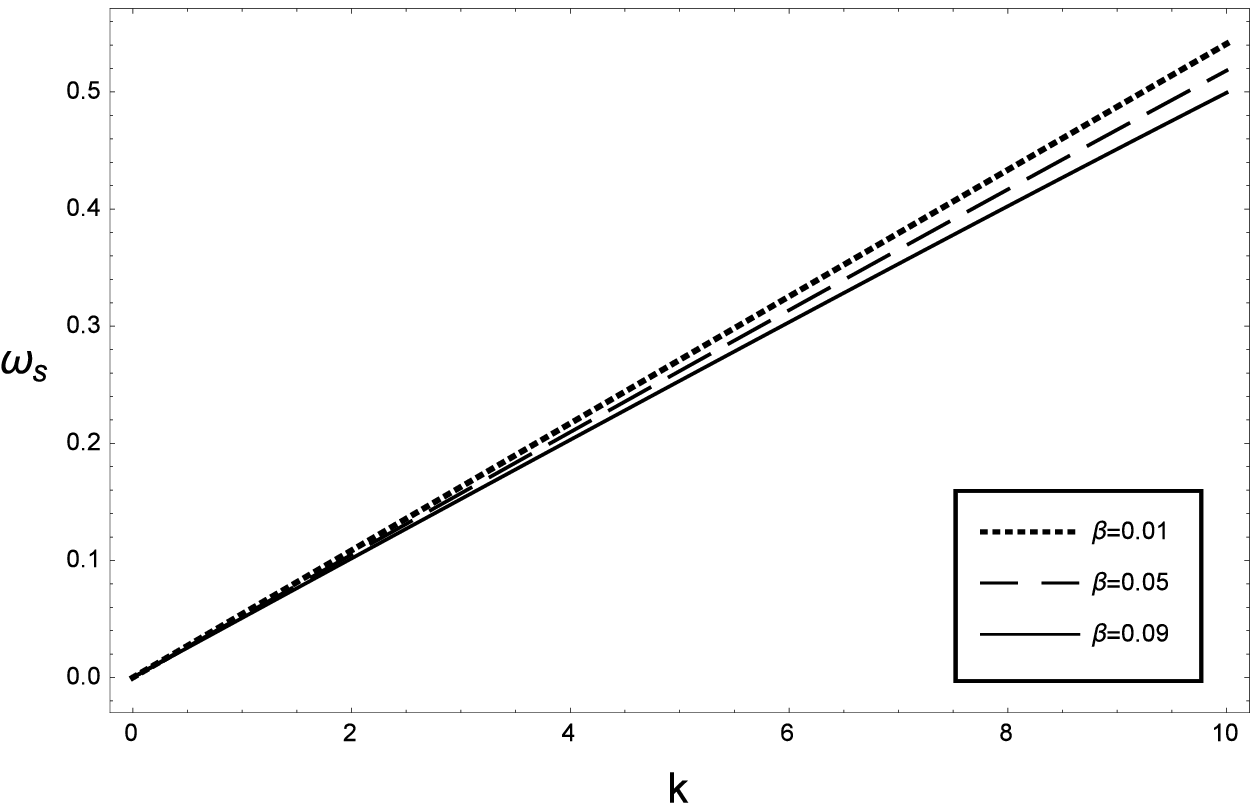}
\caption{The variation of $\omega_s$ with $k$ for different values of $\beta$;
along with $\alpha=2.0$, $\delta=0.3$, $\mu_i=0.4$, $\sigma_1=0.0001$, $\sigma_2=0.001$, and $q=1.5$.}
\label{1Fig:F2}
\end{figure}

We will use the reductive perturbation method
to derive the NLS equation to know the modulation of the DAWs. Let us consider, the stretched co-ordinate
\cite{Chowdhury2017a} as
\begin{eqnarray}
&&\hspace*{-5.5cm}\xi=\epsilon(X-V_g T),
\label{1eq:15}\\
&&\hspace*{-5.5cm}\tau=\epsilon^2T, \label{1eq:16}
\end{eqnarray}
where $V_g$ is the envelope group velocity and $\epsilon$ ($0<\epsilon<1$) is a small (real)
parameter. The dependent variables \cite{Chowdhury2017a} can be written as
\begin{eqnarray}
&&\hspace*{-1.5cm}N_1=1 +\sum_{m=1}^{\infty}\epsilon^{(m)}\sum_{l=-\infty}^
{\infty}N_{1l}^{(m)}(\xi,\tau) ~\mbox{exp}(i l \Upsilon),
\label{1eq:17}\\
&&\hspace*{-1.5cm}U_1=\sum_{m=1}^{\infty}\epsilon^{(m)}\sum_{l=-\infty}^
{\infty}U_{1l}^{(m)}(\xi,\tau) ~\mbox{exp}(i l\Upsilon),
\label{1eq:18}\\
&&\hspace*{-1.5cm}N_2=1+\sum_{m=1}^{\infty}\epsilon^{(m)}\sum_{l=-\infty}^
{\infty}N_{2l}^{(m)}(\xi,\tau) ~\mbox{exp}(i l\Upsilon),
\label{1eq:19}\\
&&\hspace*{-1.5cm}U_2=\sum_{m=1}^{\infty}\epsilon^{(m)}\sum_{l=-\infty}^
{\infty}U_{2l}^{(m)}(\xi,\tau) ~\mbox{exp}(i l\Upsilon),
\label{1eq:20}\\
&&\hspace*{-1.5cm}\phi=\sum_{m=1}^{\infty}\epsilon^{(m)}\sum_{l=-\infty}^
{\infty}\phi_{l}^{(m)}(\xi,\tau) ~\mbox{exp}(i l\Upsilon)
\label{1eq:21},
\end{eqnarray}
where $\Upsilon=k X -\omega T$ and $k$($\omega$) is the carrier wave number (frequency). The
derivative operators in the above equations are considered as follows:
\begin{eqnarray}
&&\hspace*{-4cm}\frac{\partial}{\partial T}\rightarrow\frac{\partial}{\partial T}-
\epsilon V_g \frac{\partial}{\partial\xi}+\epsilon^2\frac{\partial}{\partial\tau},\\
\label{1eq:22}
&&\hspace*{-4cm}\frac{\partial}{\partial X}\rightarrow\frac{\partial}{\partial X}+
\epsilon\frac{\partial}{\partial\xi}.
\label{1eq:23}
\end{eqnarray}
Now, by substituting Eqs. (\ref{1eq:15})$-$(\ref{1eq:23}) into Eqs. (\ref{1eq:7})$-$(\ref{1eq:10}),
and (\ref{1eq:14}), and collecting power term of $\epsilon$, the first order
approximation ($m=1$) with the first harmonic ($l=1$) provides the following relation
\begin{eqnarray}
&&\hspace*{-2.8cm}i k U_{11}^{(1)}-i\omega N_{11}^{(1)}=0,
\label{1eq:24}\\
&&\hspace*{-2.8cm}i k \lambda N_{11}^{(1)}-i\omega U_{11}^{(1)}-i k \phi_1^{(1)}=0,
\label{1eq:25}\\
&&\hspace*{-2.8cm}i k U_{21}^{(2)}-i\omega N_{21}^{(2)}=0,
\label{1eq:26}\\
&&\hspace*{-2.8cm}i k \theta N_{21}^{(1)}+i k \alpha \phi_1^{(1)}-i \omega U_{21}^{(1)}=0,
\label{1eq:27}\\
&&\hspace*{-2.8cm}\beta N_{21}^{(1)}-N_{11}^{(1)}-k^2 \phi_1^{(1)}-\gamma_1 \phi_1^{(1)}=0,
\label{1eq:28}
\end{eqnarray}
where $\lambda=3 \sigma_1$ and $\theta=3 \sigma_2$. Now, these equations detracted to
the following pattern
\begin{eqnarray}
&&\hspace*{-5.9cm}N_{11}^{(1)}=\frac{k^2}{S}\phi_1^{(1)},
\label{1eq:29}\\
&&\hspace*{-5.9cm}U_{11}^{(1)}=\frac{k \omega}{S}\phi_1^{(1)},
\label{1eq:30}\\
&&\hspace*{-5.9cm}N_{21}^{(2)}=\frac{k^2 \alpha}{A}\phi_1^{(1)},
\label{1eq:31}\\
&&\hspace*{-5.9cm}U_{21}^{(1)}=\frac{k \omega \alpha}{A}\phi_1^{(1)}
\label{1eq:32},
\end{eqnarray}
where $A=\omega^2-\theta k^2$ and $S=\lambda k^2-\omega^2$. Therefore, the dispersion
relation for the DAWs can be written as
\begin{eqnarray}
&&\hspace*{-3.8cm}\omega^2=\frac{k^2G\pm k^2 \sqrt{G^2-4HM}}{2H},
\label{1eq:33}
\end{eqnarray}
where $G=(\theta k^2+\lambda k^2+\theta \gamma_1+\lambda\gamma_1+\alpha\beta+1), H=(k^2+\gamma_1)$,
and $M=(\theta\lambda k^2+\theta\gamma_1\lambda+\theta+\alpha \beta\lambda)$. The condition
$G^2>4HM$ must be satisfied in order to obtain real and positive values of $\omega$. Normally, two
types of DA modes exist, namely, fast ($\omega_f$) and slow ($\omega_s$) DA modes according to
the positive and negative sign of the Eq. (\ref{1eq:33}). Now, we have studied the dispersion properties
by depicting $\omega$ with $k$ in Figs. \ref{1Fig:F1} and \ref{1Fig:F2}, which clearly indicates that (a) the value of $\omega_f$
increases exponentially with the increasing values of $z_2$ for fixed value of $z_1$, $n_{20}$, and $n_{10}$
although, a saturation region begins after a certain value of $k$ (as is clearly shown in Fig. \ref{1Fig:F1});
(b) on the other side, the magnitude of $\omega_s$ linearly increases with the increase of $z_2$ for the fixed
value of $z_1$, $n_{20}$, and $n_{10}$ (as is clearly shown in Fig. \ref{1Fig:F2}). It is important to
mention that in fast DA mode, both positive and negative warm dust species oscillate in phase with
electrons and ions, whereas, in slow DA mode, only one of the inertial massive dust components
oscillate in phase with electrons and ions, but the other species are in anti-phase with them
\cite{Chowdhury2017a}. Next, with the help of second-order ($m=2$ with $l=1$) equations, we obtain
the expression of $V_g$ like that
\begin{eqnarray}
&&\hspace*{-4.9cm}V_g=\frac{F_1}{2 \omega k(A^2+\alpha \beta S^2)},
\label{1eq:34}
\end{eqnarray}
\begin{eqnarray}
&&\hspace*{-2.0cm}F_1=\lambda A^2 k^2+\alpha \beta \theta k^2 S^2+\omega^2 A^2+\alpha \beta \omega^2  S^2
\nonumber\\
&&\hspace*{-1.2cm}-2 A^2 S^2-S A^2+ \alpha\beta A S^2.\nonumber \
\end{eqnarray}
Now, the second-harmonic mode of the carrier comes from nonlinear self interaction caused by
the components ($l=2$) for the second order ($m=2$) compressed equations in the following form
\begin{eqnarray}
&&\hspace*{-5.9cm}N_{12}^{(2)}=C_1|\phi_1^{(1)}|^2,
\label{1eq:35}\\
&&\hspace*{-5.9cm}U_{12}^{(2)}=C_2|\phi_1^{(1)}|^2,
\label{1eq:36}\\
&&\hspace*{-5.9cm}N_{22}^{(2)}=C_3|\phi_1^{(1)}|^2,
\label{1eq:37}\\
&&\hspace*{-5.9cm}U_{22}^{(2)}=C_4|\phi_1^{(1)}|^2,
\label{1eq:38}\\
&&\hspace*{-5.9cm}\phi_2^{(2)}=C_5|\phi_1^{(1)}|^2,
 \label{1eq:39}
\end{eqnarray}
where
\begin{eqnarray}
&&\hspace*{-0.01cm}C_1=\frac{2 C_5 k^2 S^2-(3 \omega^2 k^4+\lambda k^6)}{2 S^3},
\nonumber\\
&&\hspace*{-0.01cm}C_2=\frac{\omega C_1 S^2-\omega k^4}{k S^2},
\nonumber\\
&&\hspace*{-0.01cm}C_3=\frac{3 \alpha^2 \omega^2 k^4+\theta \alpha^2 k^6+2 \alpha C_5 A^2 k^2}{2 A^3},
\nonumber\\
&&\hspace*{-0.01cm}C_4=\frac{\omega C_3 A^2-\omega \alpha^2 k^4}{k A^2},
\nonumber\\
&&\hspace*{-0.01cm}C_5=\frac{F_2}{F_3},
\nonumber\\
&&\hspace*{-0.01cm}F_2=3\omega^2 A^3 k^4+\lambda A^3 k^6-2 \gamma_2 A^3 S^3+3 \beta \alpha^2 \omega^2 S^3 k^4
\nonumber\\
&&\hspace*{+0.8cm}+\beta \theta \alpha^2 S^3 k^6,
\nonumber\\
&&\hspace*{-0.01cm}F_3=2k^2S^2A^3+8k^2A^3S^3+2\gamma_1 A^3S^3-2 \alpha \beta A^2 k^2 S^3.
\nonumber
\end{eqnarray}
After that, we consider the expression for ($m=3,l=0$) and ($m=2,l=0$), which leads the zeroth harmonic modes. Finally, we get
\begin{eqnarray}
&&\hspace*{-5.5cm}N_{10}^{(2)}=C_6|\phi_1^{(1)}|^2,
\label{1eq:40}\\
&&\hspace*{-5.5cm}U_{10}^{(2)}=C_7|\phi_1^{(1)}|^2,
\label{1eq:41}\\
&&\hspace*{-5.5cm}N_{20}^{(2)}=C_8|\phi_1^{(1)}|^2,
\label{1eq:42}\\
&&\hspace*{-5.5cm}U_{20}^{(2)}=C_9|\phi_1^{(1)}|^2,
\label{1eq:43}\\
&&\hspace*{-5.5cm}\phi_0^{(2)}=C_{10}|\phi_1^{(1)}|^2,
\label{1eq:44}
\end{eqnarray}
where
\begin{eqnarray}
&&\hspace*{-0.7cm}C_6=\frac{2  \omega V_g k^3+\lambda k^4+ \omega^2 k^2-C_{10} S^2}{S^2 (V_g^2-\lambda)},
\nonumber\\
&&\hspace*{-0.7cm}C_7=\frac{C_6 V_g S^2-2\omega k^3}{S^2},
\nonumber\\
&&\hspace*{-0.7cm}C_8=\frac{2\omega V_g \alpha^2 k^3+\theta \alpha^2 k^4+\alpha^2 \omega^2k^2 +\alpha C_{10} A^2}{A^2 (V_g^2-\theta)},
\nonumber\\
&&\hspace*{-0.7cm}C_9=\frac{ C_8 V_g A^2-2\omega \alpha^2 k^3}{A^2},
\nonumber\\
&&\hspace*{-0.7cm}C_{10}=\frac{F_4}{F_5},
\nonumber\\
&&\hspace*{-0.7cm}F_4=(A^2 V_g^2-\theta A^2)(2\omega V_g k^3+\lambda k^4+\omega^2k^2)
\nonumber\\
&&\hspace*{+0.2cm}+(V_g^2-\lambda)(2\gamma_2 A^2 S^2 V_g^2-2\gamma_2\theta A^2 S^2)
\nonumber\\
&&\hspace*{+0.2cm}-(2\omega V_g\alpha^2k^3+\theta \alpha^2k^4+\alpha^2\omega^2 k^2)
\nonumber\\
&&\hspace*{+0.2cm}(\beta S^2V_g^2-\beta \lambda S^2 ),
\nonumber\\
&&\hspace*{-0.7cm}F_5=\alpha \beta A^2 S^2 V_g^2-\alpha \beta \lambda A^2 S^2+A^2 S^2 V_g^2-\theta A^2 S^2 
\nonumber\\
&&\hspace*{+0.2cm}-(V_g^2 -\lambda)(\gamma_1 A^2 S^2 V_g^2-\gamma_1 \theta A^2 S^2)
\nonumber.
\end{eqnarray}
Later, with the help of Eqs. (\ref{1eq:29})$-$(\ref{1eq:44}), the third harmonic ($m=3$) with ($l=1$) modes
can be accomplished. Thus, we can write the NLS equation as;
\begin{eqnarray}
&&\hspace*{-3.9cm}i\frac{\partial \Phi}{\partial \tau}+P\frac{\partial^2 \Phi}{\partial \xi^2}+Q|\Phi|^2\Phi=0,
\label{1eq:45}
\end{eqnarray}
where $\Phi=\phi_1^{(1)}$ for simplicity and $P$ ($Q$) is the dispersion (nonlinear) coefficient, and is written by
\begin{eqnarray}
&&\hspace*{-4.2cm}P=\frac{F_6}{2 \omega A S k^2 (A^2+\alpha \beta S^2)},
\nonumber\
\end{eqnarray}
and
\begin{eqnarray}
&&\hspace*{-4.7cm}Q=\frac{F_7}{2 \omega k^2 (A^2+\alpha \beta S^2)},
\nonumber\
\end{eqnarray}
where
\begin{eqnarray}
&&\hspace*{-0.45cm}F_6=(\omega V_g A^3-\lambda k A^3 )(\lambda k^3-2\omega V_g k^2 +k \omega^2 -k S)
\nonumber\\
&&\hspace*{+0.4cm}+(k V_g A^3-\omega A^3 )(\lambda \omega k^2-2k V_g \omega^2 +\omega^3-k S V_g)
\nonumber\\
&&\hspace*{+0.4cm}-(\theta k^3-2 \omega V_g k^2+k \omega^2+A k)(\alpha \beta \omega V_g S^3 
\nonumber\\
&&\hspace*{+0.4cm}-\alpha \beta \theta k S^3)-A^3 S^3-(\alpha \beta k V_g S^3 -\alpha \beta \omega S^3)
\nonumber\\
&&\hspace*{+0.4cm}(\theta \omega k^2-2 k V_g \omega^2+\omega^3+ A k V_g),
\nonumber\
\end{eqnarray}
\begin{eqnarray}
&&\hspace*{-1.2cm}F_7=2 \gamma_2 C_5 A^2 S^2 +2 \gamma_2 C_{10} A^2 S^2+3 \gamma_3 A^2 S^2
\nonumber\\
&&\hspace*{-0.5cm}-2 \omega C_2 A^2 k^3 - 2\omega C_7 A^2 k^3-2 \alpha \beta \omega C_4 S^2 k^3
\nonumber\\
&&\hspace*{-0.5cm} -2 \alpha \beta \omega C_9 S^2 k^3-C_1 \omega^2 A^2 k^2- C_6 \omega^2 A^2 k^2
\nonumber\\
&&\hspace*{-0.5cm}-\lambda C_1 A^2 k^4-\lambda C_6 A^2 k^4- \alpha \beta C_3 \omega^2 k^2 S^2 
\nonumber\\
&&\hspace*{-0.5cm}-\alpha \beta C_8 \omega^2 k^2 S^2 -\alpha \beta \theta C_3 S^2 k^4-\alpha \beta \theta C_8 S^2 k^4.
\nonumber
\end{eqnarray}
\begin{figure}[t!]
\centering
\includegraphics[width=70mm]{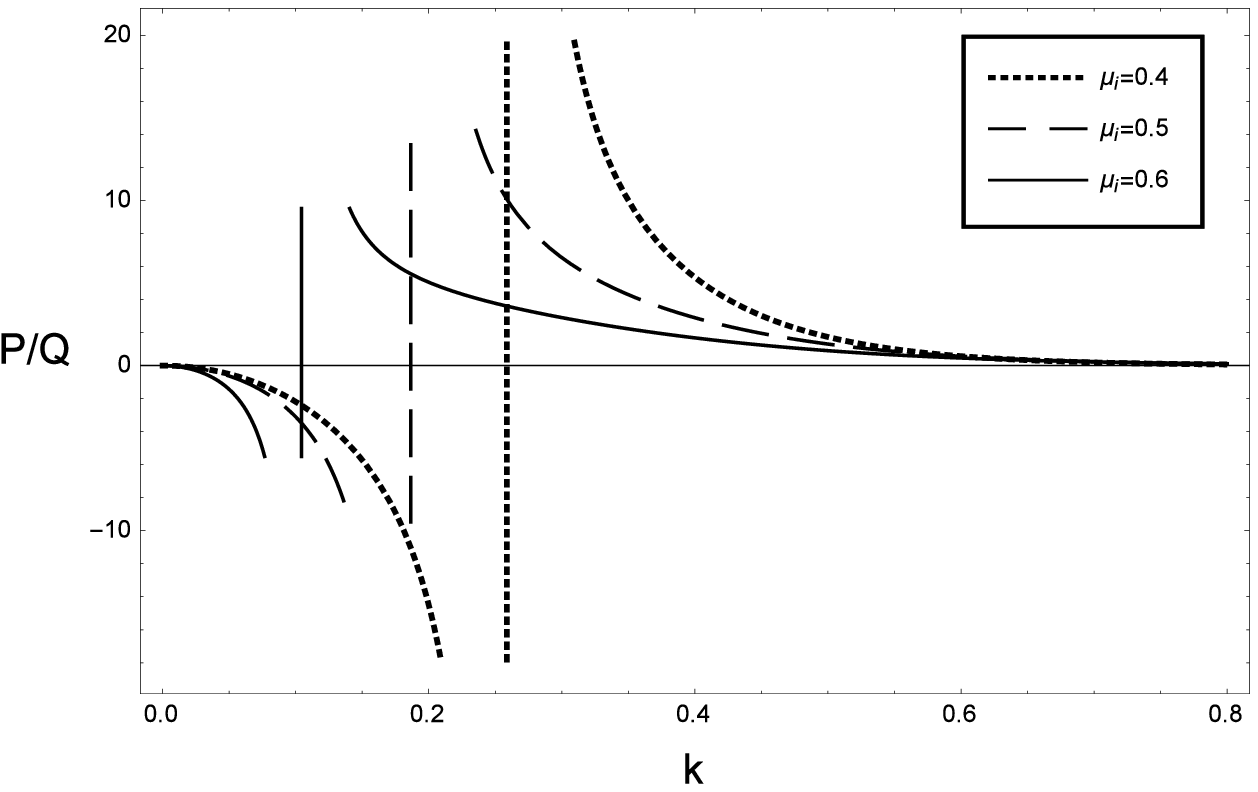}
\caption{The variation of $P/Q$ with $k$ for different values of $\mu_i$;
along with $\alpha=2.0$, $\beta=0.07$, $\delta=0.3$,  $\sigma_1=0.0001$, $\sigma_2=0.001$, $\omega_f$, and $q=1.5$.}
\label{1Fig:F3}
\vspace{0.8cm}
\includegraphics[width=70mm]{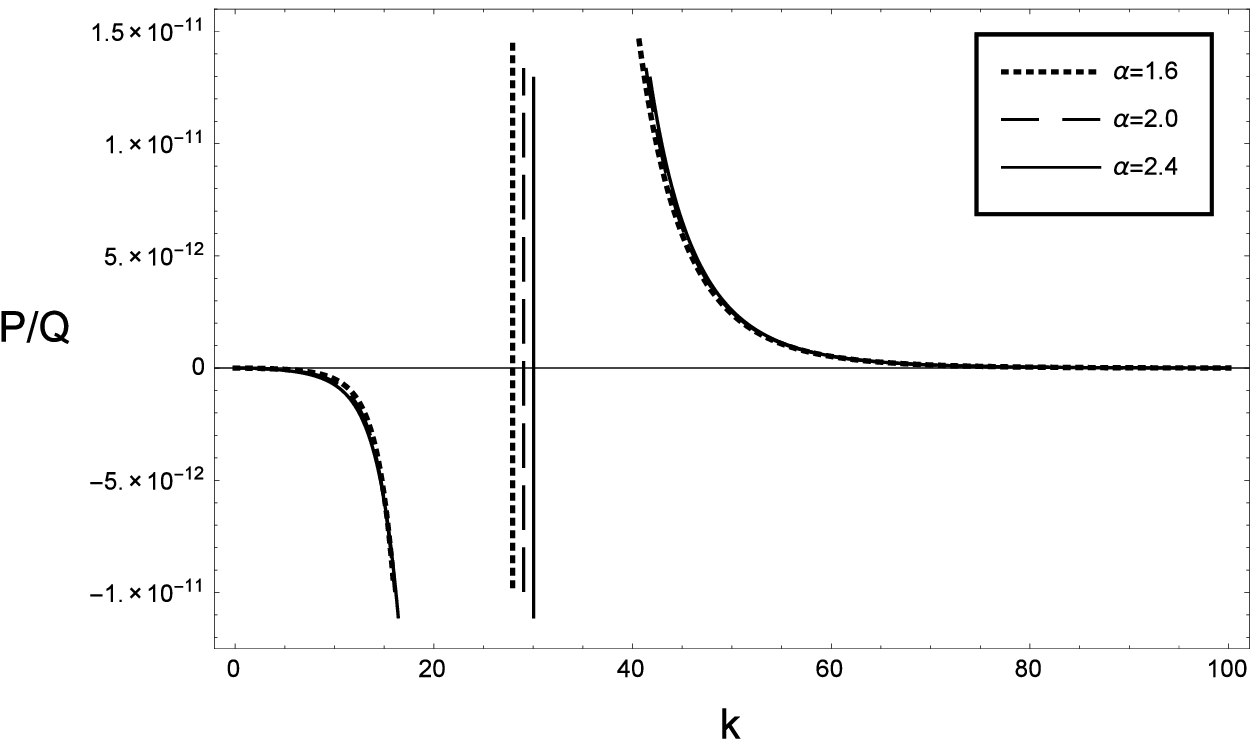}
\caption{The variation of $P/Q$ with $k$ for different values of $\alpha$;
along with $\beta=0.07$, $\delta=0.3$, $\mu_i=0.4$, $\sigma_1=0.0001$, $\sigma_2=0.001$, $\omega_s$, and $q=1.5$.}
\label{1Fig:F4}
\end{figure}
\begin{figure}[t!]
\centering
\includegraphics[width=70mm]{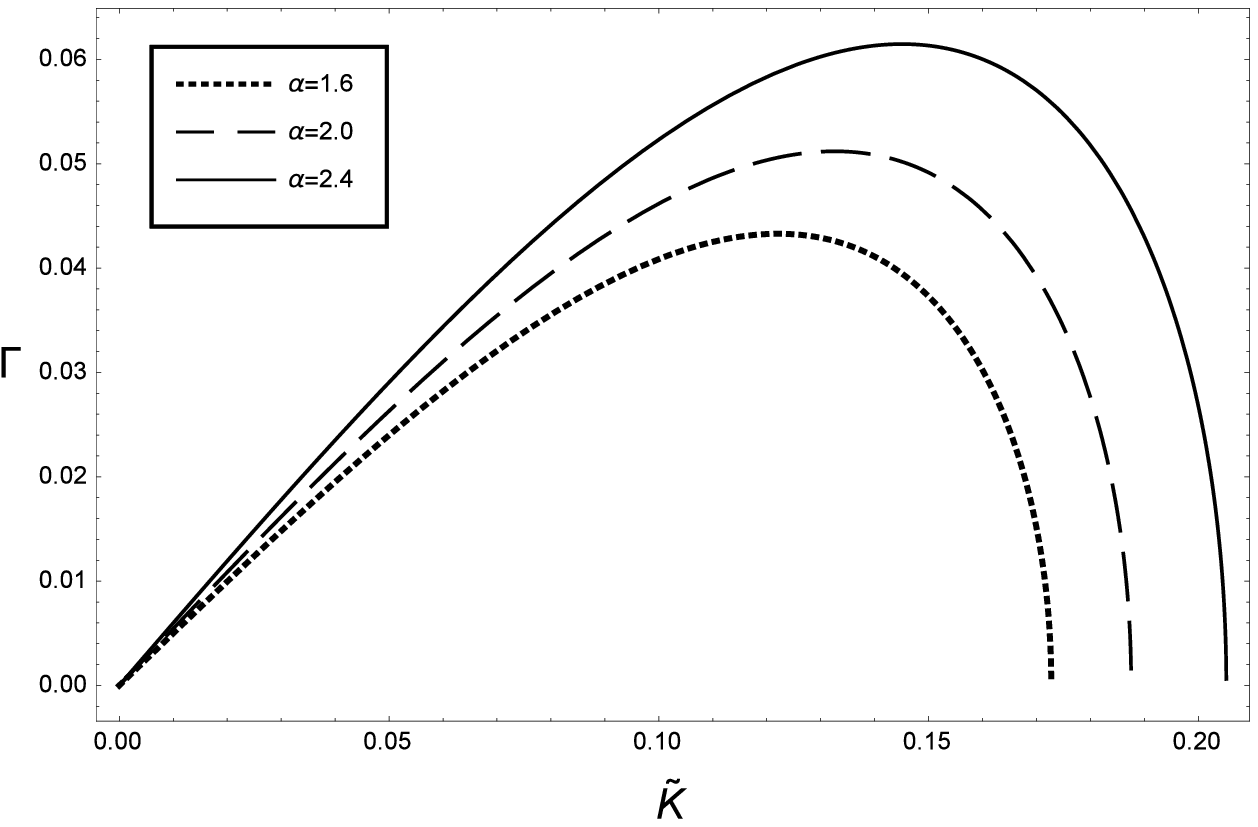}
\caption{The variation of MI growth rate $(\Gamma)$ with $\tilde{k}$ for different values of $\alpha$;
along with $\beta=0.07$, $\delta=0.3$, $\mu_i=0.4$, $\sigma_1=0.0001$, $\sigma_2=0.001$, $\omega_f$, $q=1.0$, $\Phi_0=0.5$, and $k=0.3$.}
\label{1Fig:F5}
\vspace{0.8cm}
\includegraphics[width=70mm]{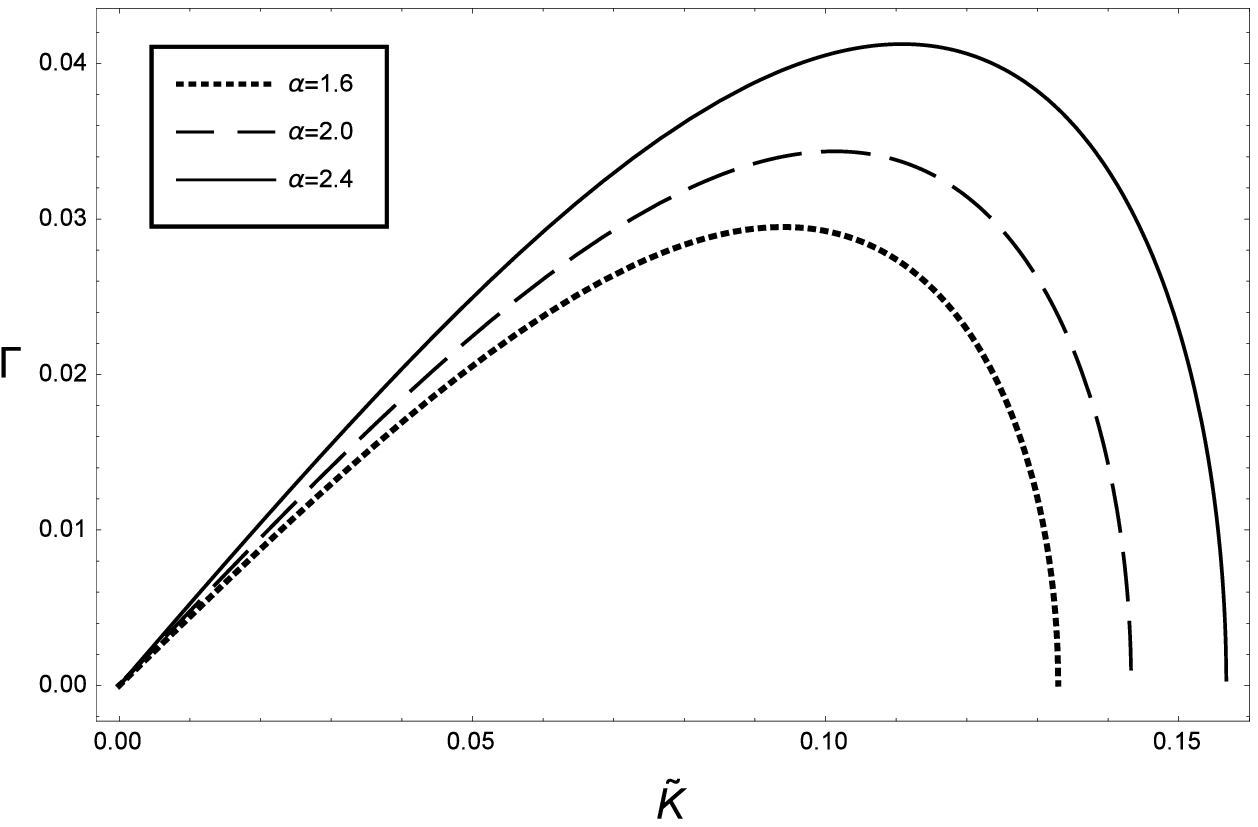}
\caption{The variation of MI growth rate $(\Gamma)$ with $\tilde{k}$ for different values of $\alpha$;
along with $\beta=0.07$, $\delta=0.3$, $\mu_i=0.4$, $\sigma_1=0.0001$, $\sigma_2=0.001$, $\omega_f$, $q=1.5$, $\Phi_0=0.5$, and $k=0.3$.}
\label{1Fig:F6}
\vspace{0.8cm}
\includegraphics[width=70mm]{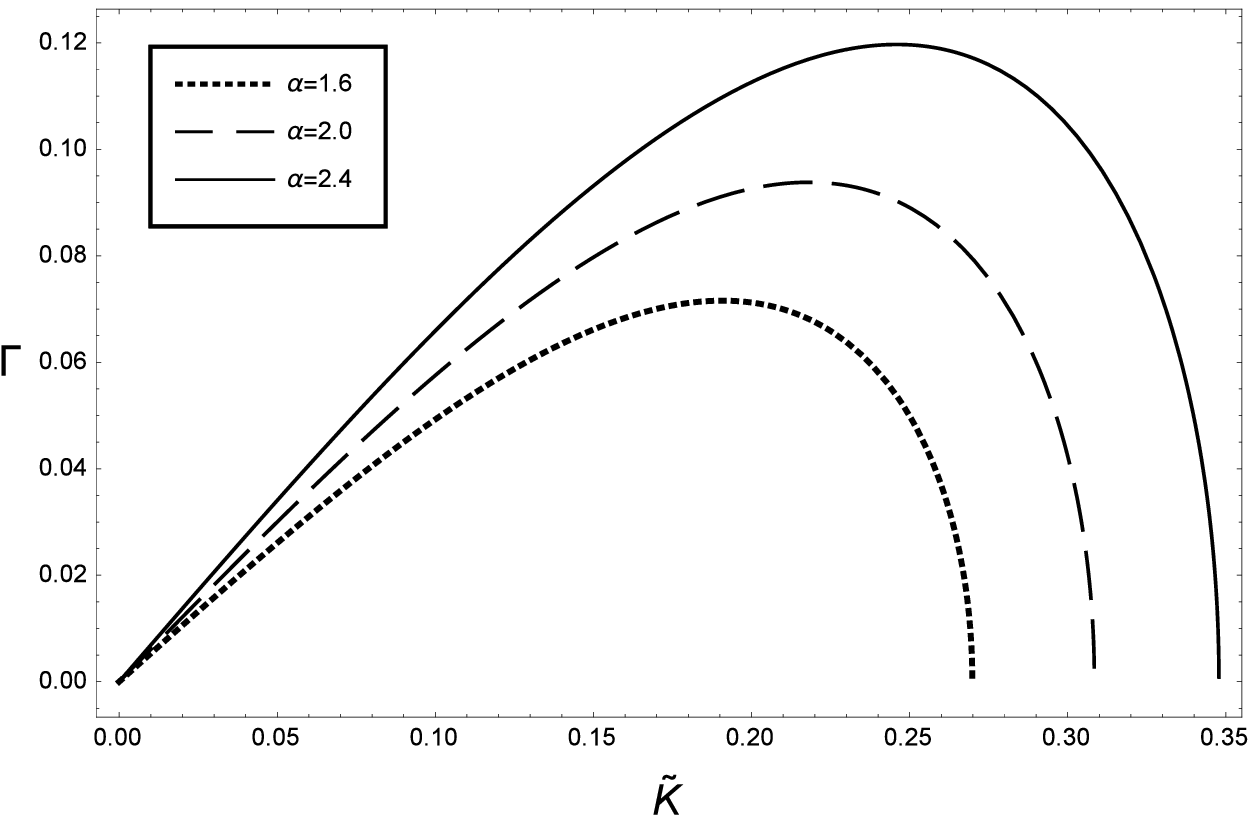}
\caption{The variation of MI growth rate $(\Gamma)$ with $\tilde{k}$ for different values of $\alpha$;
along with $\beta=0.07$, $\delta=0.3$, $\mu_i=0.4$, $\sigma_1=0.0001$, $\sigma_2=0.001$, $\omega_f$, $q=-0.6$, $\Phi_0=0.5$, and $k=0.3$.}
\label{1Fig:F7}
\end{figure}
\begin{figure}[t!]
\centering
\includegraphics[width=70mm]{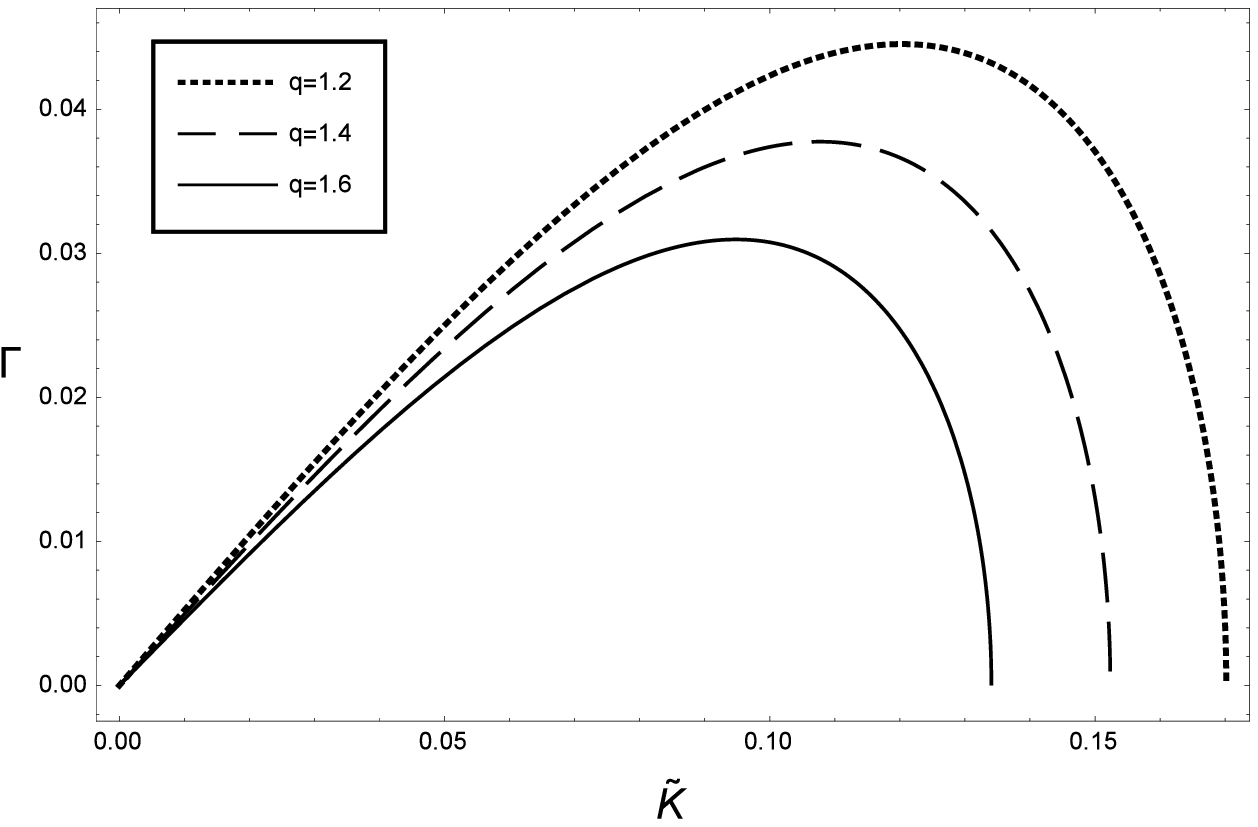}
\caption{The variation of MI growth rate $(\Gamma)$ with $\tilde{k}$ for $q$ (within sub-extensive range);
along with $\beta=0.07$, $\delta=0.3$, $\mu_i=0.4$, $\sigma_1=0.0001$, $\sigma_2=0.001$, $\omega_f$, $\Phi_0=0.5$, and $k=0.3$.}
\label{1Fig:F8} \vspace{0.8cm}
\includegraphics[width=70mm]{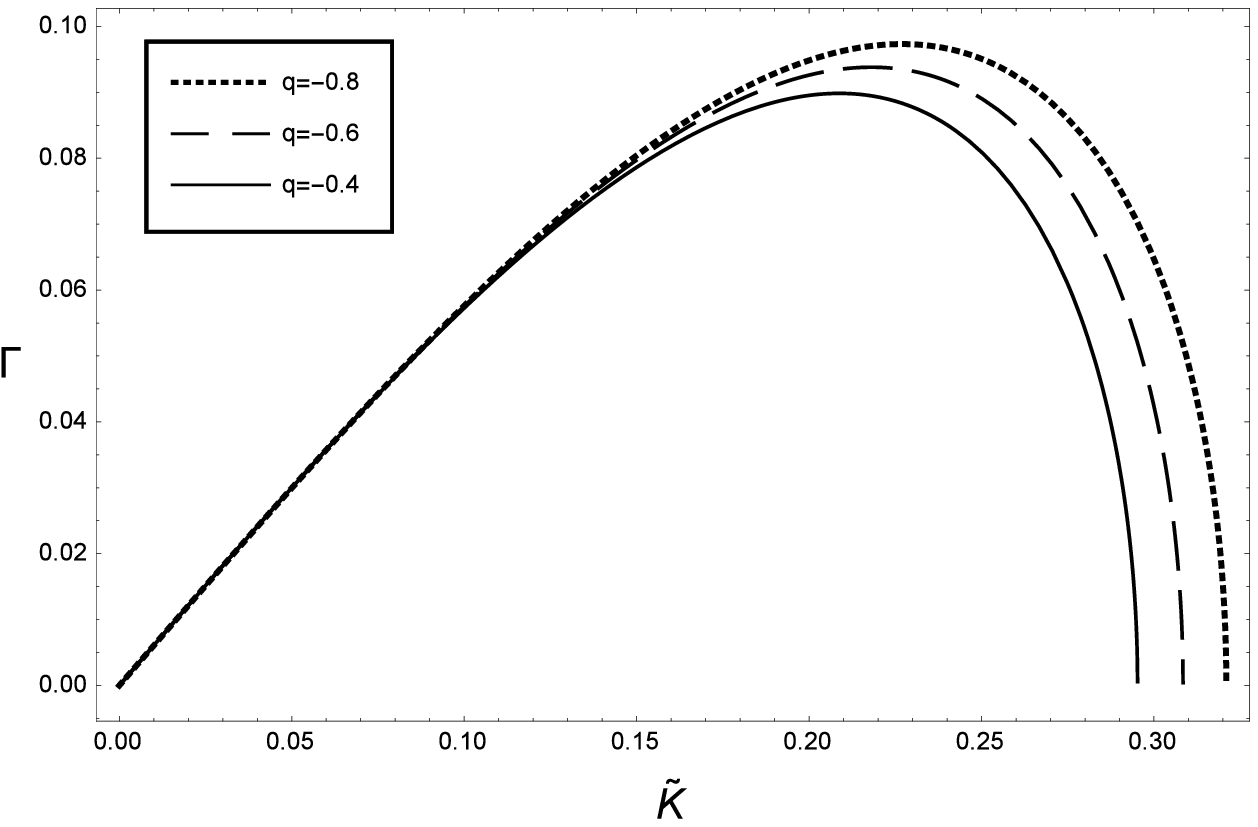}
\caption{The variation of MI growth rate $(\Gamma)$ with $\tilde{k}$ for $q$ (within super-extensive range);
along with $\beta=0.07$, $\delta=0.3$, $\mu_i=0.4$, $\sigma_1=0.0001$, $\sigma_2=0.001$, $\omega_f$, $\Phi_0=0.5$, and $k=0.3$.}
\label{1Fig:F9}
\end{figure}
\begin{figure}[t!]
\centering
\includegraphics[width=70mm]{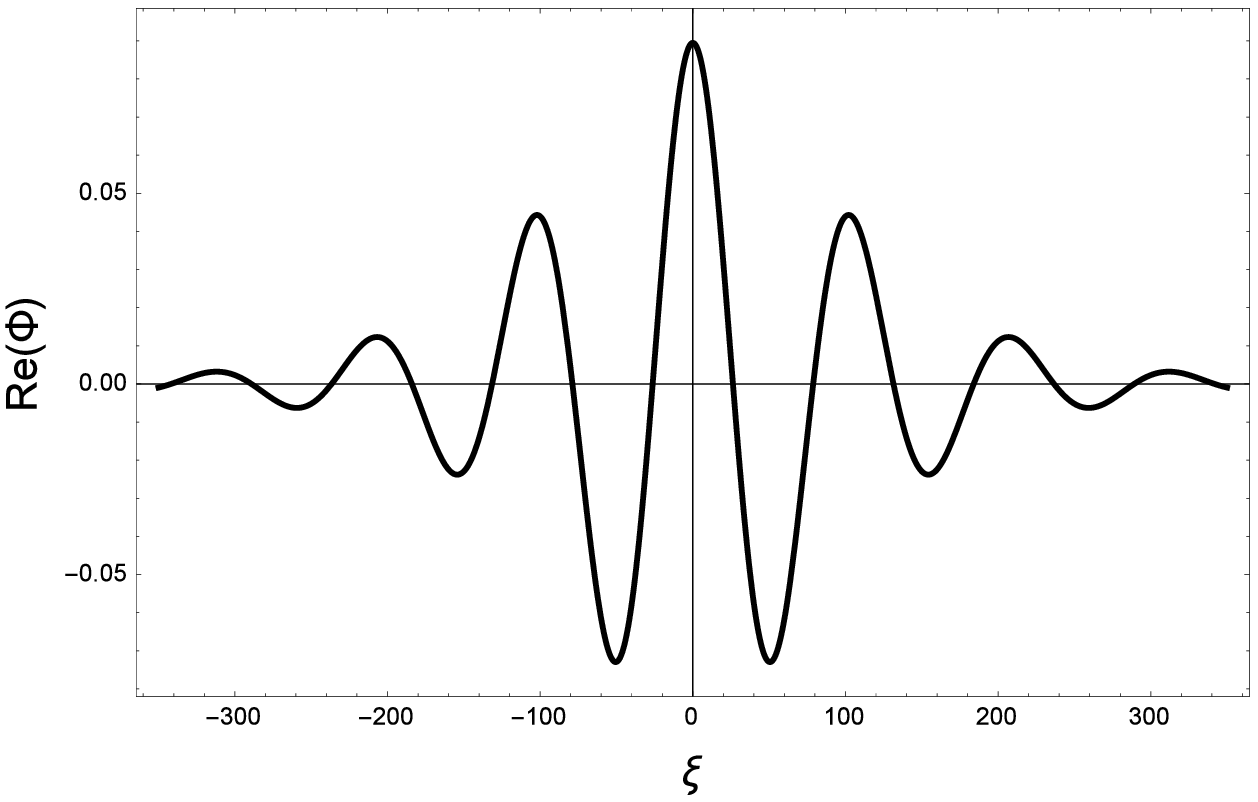}
\caption{The variation of $Re(\Phi)$ with $\xi$ for bright envelope solitons;
along with $\alpha=2.0$, $\beta=0.07$, $\delta=0.3$, $\mu_i=0.4$,  $\sigma_1=0.0001$, $\sigma_2=0.001$,
$\tau=0$, $\psi_0=0.008$, $\omega_f$, $q=1.5$, $k=0.3$, $\Omega_0=0.4$, and $U=0.4$.}
\label{1Fig:F10}
\vspace{0.8cm}
\includegraphics[width=70mm]{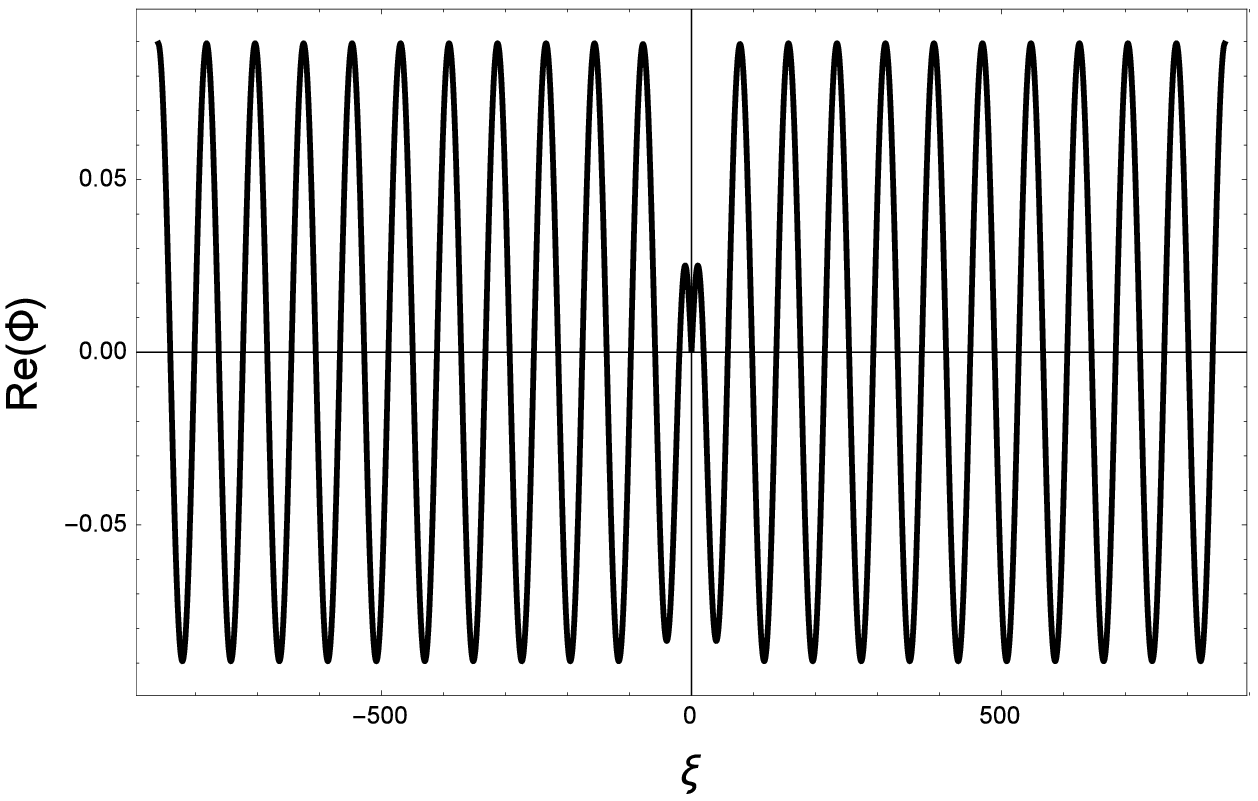}
\caption{The variation of $Re(\Phi)$ with $\xi$ for dark envelope solitons;
along with $\alpha=2.0$, $\beta=0.07$, $\delta=0.3$, $\mu_i=0.4$,  $\sigma_1=0.0001$, $\sigma_2=0.001$,
$\tau=0$, $\psi_0=0.008$, $\omega_f$, $q=1.5$, $k=0.1$, $\Omega_0=0.4$, and $U=0.4$.}
\label{1Fig:F11}
\end{figure}
\section{Stability of DAWs}
\label{1sec:Stability of DAWs} To study the MI of DAWs, we consider the linear solution of the
NLS equation (\ref{1eq:45}) in the form $\Phi=\hat{\Phi}e^{iQ|\hat{\Phi}|^2 \tau}+c.c$ (c.c denotes the
complex conjugate), where $\hat{\Phi}=\hat{\Phi}_0+\epsilon \hat{\Phi}_1$ and
$\hat{\Phi}_1=\hat{\Phi}_{1,0}e^{i(\tilde{k}\xi-\tilde{\omega}\tau)}$+c.c (the perturbed wave number $\tilde{k}$
and the frequency $\tilde{\omega}$ are different from $k$ and $\omega$). Now, by substituting these
values in Eq. (\ref{1eq:45}) the following nonlinear dispersion relation can be obtained as
\cite{Sultana2011,Kourakis2005,Schamel2002,Fedele2002}
\begin{eqnarray}
&&\hspace*{-4cm}\tilde{\omega}^2=P^2\tilde{k}^2 \left(\tilde{k}^2-\frac{2|\hat{\Phi}_0|^2}{P/Q}\right).
\label{1eq:46}
\end{eqnarray}
It is clear from Eq. (\ref{1eq:46}) that the DAWs will be modulationally stable (unstable) in the
range of values of $\tilde{k}$ in which $P/Q$ is negative (positive). When
$P/Q\rightarrow\pm\infty$, the corresponding value of $k$ ($=k_c$) is called the critical or threshold
wave number ($k_c$) for the onset of MI. The variation of $P/Q$ with $k$ for $\mu_i$ and
$\alpha$ are shown in Figs. \ref{1Fig:F3} and \ref{1Fig:F4}, respectively, which clearly indicate that
(a) the value of $k_c$ increases with the increase of $n_{i0}$ for fixed value of $z_1$ and $n_{10}$;
(b) on the other hand, $k_c$ value decreases with the increase of $m_2$ for fixed value of $m_1$, $z_2$,
and $z_1$. It also can be written from Eq. (\ref{1eq:46}) that the growth rate ($\Gamma$) of the
modulationally unstable region for the DAWs [when $P/Q>0$ and $\tilde{k}<\tilde{k}_c=(2 Q|\hat{\Phi}_0|^2/P)^{1/2}$] as
\begin{eqnarray}
&&\hspace*{-4.5cm}\Gamma=|P|\tilde{k}^2\left(\frac{\tilde{k}^2_c}{\tilde{k}^2}-1\right)^{1/2}.
\label{1eq:47}
\end{eqnarray}
Moreover, we have graphically shown how the $\Gamma$ varies with $\tilde{k}$ of different values of
$\alpha$ and $q$ in  Figs. \ref{1Fig:F5}$-$\ref{1Fig:F9}.

It is obvious from Figs. \ref{1Fig:F5}$-$\ref{1Fig:F7} that (a) the maximum value of $\Gamma$ increases
with the increase of the values of $z_2$ for fixed value of $q=1$ (Maxwellian limit), $z_1$, $m_1$, and
$m_2$ (as is clearly shown in Fig. \ref{1Fig:F5}); (b) the magnitude of $\Gamma$ also increases with
the increase of $z_2$ for fixed value of $q=1.5$ (subextensive range), $z_1$, $m_1$, and $m_2$ (as is
clearly shown in Fig. \ref{1Fig:F6}); (c) Moreover, $\Gamma$ increases with the increase of $z_2$ for
fixed value of $q=-0.6$ (super-extensive range), $z_1$, $m_1$, and $m_2$ (as clearly shown in Fig.
\ref{1Fig:F7}). The physics of this result is that, since the nonlinearity increases with the increase
of the values of $\alpha$, the growth rate of DAWs increase.

It can be observed from Figs. \ref{1Fig:F8} and \ref{1Fig:F9} that the maximum value of $\Gamma$ increases
(decreases) with the decrease (increase) of the values of $q>1$ ($q<1$) (as is clearly shown in Fig.
\ref{1Fig:F8} and \ref{1Fig:F9}). It can be concluded here that the variation of $\Gamma$ with respect to $\tilde{k}$ is
independent on the sign of the $q$.
\section{Envelope solitons}
\label{1sec:Envelope solitons} There are two types of envelope solitonic solutions exist, namely, bright
and dark envelope solitons, depending on the sign of the coefficients $P$ and $Q$.
\subsection{Bright envelope solitons}
When $P/Q>0$, the expression of the bright envelope solitonic solution of Eq. (\ref{1eq:45}) can be
written as \cite{Sultana2011,Kourakis2005,Schamel2002,Fedele2002}
\begin{eqnarray}
&&\hspace*{-1.5cm}\Phi(\xi,\tau)=\left[\psi_0~\mbox{sech}^2 \left(\frac{\xi-U\tau}{W}\right)\right]^{1/2}
\nonumber\\
&&\hspace*{-0.2cm}\times \exp \left[\frac{i}{2P}\left\{U\xi+\left(\Omega_0-\frac{U^2}{2}\right)\tau \right\}\right],
\label{1eq:48}
\end{eqnarray}
where $\psi_0$ indicates the envelope amplitude, $U$ is the travelling speed of the localized pulse, $W$ is
the pulse width, which can be written as $W=(2P\psi_0/Q)^{1/2}$, and $\Omega_0$ is the
oscillating frequency for $U=0$. The bright envelope soliton is depicted in Fig. \ref{1Fig:F10}.
The amplitude of the bright envelope solitons remains constant but the width of the bright envelope solitons  increases as
we increase the value of the $q$-distributed electron temperature $T_e$, for fixed value of ion temperature $T_i$ (via $\delta$).
\subsection{Dark envelope solitons}
As we know before that the condition for dark envelope soliton is $P/Q<0$. So, the dark envelope soliton
solution of Eq. (\ref{1eq:45}) can be  written as
\cite{Sultana2011,Kourakis2005,Schamel2002,Fedele2002}
\begin{eqnarray}
&&\hspace*{-1.5cm}\Phi(\xi,\tau)=\left[\psi_0~\mbox{tanh}^2 \left(\frac{\xi-U\tau}{W}\right)\right]^{1/2}
\nonumber\\
&&\hspace*{-0.2cm}\times \exp \left[\frac{i}{2P}\left\{U\xi-\left(\frac{U^2}{2}-2 P Q \psi_0\right)\tau \right\}\right].
\label{1eq:49}
\end{eqnarray}
The dark envelope soliton, obtained from Eq. (\ref{1eq:49}) is shown in Fig. \ref{1Fig:F11}.
\section{Discussion}
\label{1sec:Discussion} We have studied an unmagnetized realistic space dusty plasma system consists
 of non-extensive $q$-distributed electrons, isothermal ions, positively charged warm dust particles as
well as negatively charged warm dust particles. The reductive perturbation method is used to derive
the NLS equation. The results that have been found from our investigation can be summarized as follows
\begin{enumerate}
\item{The fast DA mode increases exponentially with the increasing values of $z_2$ for fixed value
of $z_1$, $n_{20}$ (via $\beta$), and $n_{10}$. On the other hand, the slow DA mode linearly increases with the increase
of $z_2$ for the fixed value of $z_1$, $n_{20}$, and $n_{10}$ (via $\beta$).}
\item{The DAWs is modulationally stable (unstable) in the range of values of $k$ in which the ratio $P/Q$
is $P/Q<0$ ($P/Q>0$).}
\item{The value of $k_c$ increases with the increase of $n_{i0}$ for fixed value of $z_1$ and $n_{10}$ (via $\mu_i$).
On the other hand, $k_c$ value decreases with the increase of $m_2$ for fixed value of $m_1$, $z_2$, and $z_1$ (via $\alpha$).}
\item{The value of $\Gamma$ increases with the decrease of $z_2$ for fixed value of $z_1$, $m_1$, $m_2$, and $q$
(where the values of $q$ lies in the range of $q>1$, $q=1$, and $q<1$, respectively).}
\item{The maximum value of $\Gamma$ increases
(decreases) with the decrease (increase) of the values of $q>1$ ($q<1$). So, the growth rate is
independent on the sign of the $q$.}
\item{The amplitude of the bright envelope solitons remains constant but the width of the bright envelope solitons increases as
we increase the value of the $q$-distributed electron temperature $T_e$, for fixed value of ion temperature $T_i$ (via $\delta$).}
\end{enumerate}
The results of our present investigation will be useful in understanding the nonlinear phenomena both
in space (e.g. Jupiters magnetosphere
\cite{Horányi1996b,Horányi1993a,Mamun2008c,Chowdhury2017a}, upper mesosphere \cite{Havnes1996}, comets
tails \cite{Horányi1996b,Mendis1991b}, etc.) and laboratory (viz. direct current and
radio-frequency discharges, plasma processing reactors, fusion plasma devices \cite{Shukla2002}, solid-fuel
combustion products \cite{Shukla2002}, etc.)  plasma system containing nonextensive
$q$-distributed electrons, isothermal ions, negatively and positively charged warm dust.

\end{document}